\documentclass[12pt]{article}
\usepackage{epsfig}
\usepackage{graphicx}
\usepackage{a4}
\usepackage{latexsym}
\usepackage{cite}

\usepackage{pslatex}
\usepackage[latin1]{inputenc}
\usepackage[T1]{fontenc}
\newcommand{\ep}{\mbox{$\varepsilon$}}
\newcommand{\Pzz}{P_{II}^{(0)}(z)}
\newcommand{\Poz}{\mbox{$P_{II}^{(1)}(z)$}}
\newcommand{\Ptz}{\mbox{$P_{II}^{(2)}(z)$}}
\newcommand{\Pthz}{\mbox{$P_{II}^{(3)}(z)$}}

\begin{document}
\setlength{\parskip}{0.2cm}
\setlength{\baselineskip}{0.55cm}
%-----------------------------
\begin{titlepage}
\begin{flushright} 
\hfill {\tt hep-ph/0512249}
\\
\hfill {\tt HRI-12/2005}
\end{flushright} 
\vspace{5mm} 
\begin{center} 
{\Large \bf 
%\boldmath 
On Sudakov and Soft resummations in QCD}\\
\end{center}

\vspace{10pt} 
\begin{center} 
{\bf 
V. Ravindran
%\footnote{ravindra@mri.ernet.in},   
%\footnote{neerven@lorentz.liedenuniv.nl}
}\\ 
\end{center}
\begin{center} 
{\it 
Harish-Chandra Research Institute, 
 Chhatnag Road, Jhunsi, Allahabad, India.\\
} 
\end{center}
 
\vspace{10pt} 
\begin{center}
{\bf ABSTRACT} 
\end{center} 
In this article we extract soft distribution functions 
for Drell-Yan and Higgs production processes 
using mass factorisation theorem and the perturbative results
that are known upto three loop level.  We find that 
they are maximally non-abelien.  We show that these functions
satisfy Sudakov type integro differential equations.      
The formal solutions to such equations and also to the mass factorisation
kernel upto four loop level are presented.  
Using the soft distribution function extracted from Drell-Yan
production, we show how the soft plus virtual cross section 
for the Higgs production can be obtained.  We determine 
the threshold resummation exponents upto three loop using
the soft distribution function.  

\vskip12pt 
\vskip 0.3 cm
%\noindent PACS numbers:.

%\setcounter{footnote}{0} 
%\renewcommand{\thefootnote}{\arabic{footnote}} 
%\newcommand{\xo}{\mbox{$x_1^0$}}
%\newcommand{\xt}{\mbox{$x_2^0$}}
%\newcommand{\ep}{\mbox{$\epsilon$}}
%\newcommand{\Pzz}{P_{II}^{(0)}(z)}
%\newcommand{\Poz}{\mbox{$P_{II}^{(1)}(z)$}}
%\newcommand{\Ptz}{\mbox{$P_{II}^{(2)}(z)$}}
%\newcommand{\Pthz}{\mbox{$P_{II}^{(3)}(z)$}}
%end{document} 
\end{titlepage}
The Drell-Yan(DY) production of di-leptons and Higgs boson 
production play crucial role in the hadronic colliders.  
The di-lepton production
can not only serve as a luminosity monitor but also 
provide vital information on physics beyond standard model 
at present collider Tevatron at Fermi-Lab and future Large Hadron Collider
(LHC) which is going to be up at CERN in few years.
Higgs production at such colliders will establish the
Standard Model(SM) as well as beyond SM Higgs \cite{Djouadi:2005gi,Djouadi:2005gj}.
From the theoretical side, the DY production of di-leptons and 
Higgs boson production are known upto Next to Next to leading order(NNLO) 
level in QCD.   
For DY at NLO level, see \cite{Altarelli:1978id}
and for the Higgs production at NLO level,
see \cite{Dawson:1990zj,Djouadi:1991tk,Spira:1995rr}.  
The NNLO contribution to DY can be found in 
\cite{Matsuura:1987wt,Matsuura:1988sm,Hamberg:1990np}.
Beyond NLO, the Higgs production cross sections are known only
in the large top quark mass limit.
For the NNLO soft plus virtual part of the Higgs production, see 
\cite{Harlander:2001is,Catani:2001ic} and the full 
NNLO for the Higgs production can be found in 
\cite{Harlander:2002wh,Anastasiou:2002yz,Ravindran:2003um}.
Apart from these fixed order results, the resummation programs 
for the threshold corrections to 
both DY and Higgs productions have also been very successful
\cite{Sterman:1986aj,Catani:1989ne}.
For next to next to leading logarithmic (NNLL) resummation, see 
\cite{Vogt:2000ci,Catani:2003zt}.
Due to  several important results at three loop level 
that are available in recent times
%\cite{Moch:2004pa,Vogt:2004mw,Moch:2005id,Moch:2005tm,Moch:2005ba,Blumlein:2004xt},
\cite{Moch:2004pa}-\cite{Blumlein:2004xt},
the resummation upto $N^3LL$ has also become reality \cite{Moch:2005ky,
Laenen:2005uz,Idilbi:2005ni}. 

With all these new results in both fixed order as well as 
resummed calculations,  one is now able to unravel the 
interesting structures in the perturbative results (for example: 
\cite{Blumlein:2000wh,Blumlein:2005im,Dokshitzer:2005bf}).
Along this line, 
in this paper, we extract the 
soft distribution functions of Drell-Yan and Higgs production
cross sections in perturbative QCD and show
that they do not depend on the process under consideration.  
By that we mean that the soft distribution function of 
Drell-Yan production can be got entirely
from the Higgs production by a simple multiplication of the colour 
factor $C_F/C_A$.  We prove this for the pole parts upto three loop level 
and for the finite part we could show only to those terms that are not
proportional to $\delta(1-z)$ because 
the three loop finite part proportional to $\delta(1-z)$ 
is not available yet and 
can be obtained only from the explicit fixed order computation
of bremsstrahlung contribution.
The extraction of the soft distribution function is achieved 
with the help of mass factorisation 
theorem supplemented by the recent developments in the computation 
of three loop anomalous dimensions, three loop form factors of 
quark and gluon operators and two loop bremsstrahlung contributions 
to Drell-Yan and Higgs productions.  We discuss the consequences
of our observation in the determination of soft plus virtual cross sections
and the threshold resummation exponents.  A brief account on the
soft and jet distribution functions and the
resummation exponents relevant for deep inelastic scattering
(DIS) is given.

We start by writing the partonic cross section as
\begin{eqnarray}
\hspace{-1cm} 
\hat \sigma^{sv}_I(z,q^2,\mu_R^2)&=&
\Big(Z^I(\hat a_s,\mu_R^2,\mu^2)\Big)^2~
               |\hat F^I\left(\hat a_s,Q^2,\mu^2\right)|^2~ 
%\nonumber\\[2ex]
         \delta(1-z)\otimes {\cal C} e^{\displaystyle{2 ~
   \Phi^I\left(\hat a_s, q^2,\mu^2,z\right)}},
\nonumber\\[2ex]
&&
\quad \quad \quad
\hspace{9cm} 
I=q,g
\end{eqnarray}
with the normalisation, $\hat \sigma^{sv}_{I,born}=\delta(1-z)$.
The symbol $sv$ means that we restrict to only the
soft and virtual contributions to the partonic cross sections 
$\hat \sigma^{sv}_I$.
In the above equation we have introduced a "${\cal C}$ ordered exponential"
which has the following expansion:
\begin{eqnarray}
{\cal C}e^{\displaystyle f(z) }= \delta(1-z)  + {1 \over 1!} f(z)
 +{1 \over 2!} f(z) \otimes f(z) + {1 \over 3!} f(z) \otimes f(z) \otimes f(z) 
+ \cdot \cdot \cdot
\end{eqnarray}
The function $f(z)$ is a distribution of the kind $\delta(1-z)$ 
and ${\cal D}_i$, where
\begin{eqnarray}
{\cal D}_i=\Bigg[{\ln^i(1-z) \over (1-z)}\Bigg]_+
\quad \quad \quad i=0,1,\cdot\cdot\cdot
\end{eqnarray}
and the symbol $\otimes$ means the Mellin convolution.
The letters $q$ and $g$ stand for Drell-Yan(DY) and Higgs(H) productions 
respectively.  
$q^2$($=-Q^2$) is the invariant mass of the final state 
(di-lepton pair in the case of DY and single Higgs boson 
for the Higgs production).  
$z$ is the scaling variable defined as the ratio of $q^2$ over $\hat s$, 
where $\hat s$ is the center of mass of the partonic system. 
$F^I(\hat a_s,Q^2,\mu^2)$ are the form factors
that enter in the Drell-Yan(for $I=q$) and Higgs(for $I=g$) 
production cross sections.  The functions $\Phi^I(\hat a_s,q^2,\mu^2,z)$ are
called the soft distribution functions. 
The unrenormalised(bare) strong coupling constant
$\hat a_s$ is defined as
\begin{eqnarray}
\hat a_s={\hat g^2_s \over 16 \pi^2}
\end{eqnarray}
where $\hat g_s$ is the strong coupling constant which is dimensionless in
$n=4+\ep$, with $n$ being the number of space time dimensions.  
The scale $\mu$ comes from
the dimensional regularisation in order to make the bare coupling constant $\hat g_s$
dimensionless in $n$ dimensions.

The bare coupling constant $\hat a_s$ is related to renormalised one by
the following relation:
\begin{eqnarray}
S_{\ep} \hat a_s = Z(\mu_R^2) a_s(\mu_R^2) \left(\mu^2 \over \mu_R^2\right)^{\ep \over 2}
\label{renas}
\end{eqnarray}
The scale $\mu_R$ is the 
renormalisation scale at which the renormalised strong coupling constant
$a_s(\mu_R)$ is defined.   
\begin{eqnarray}
S_{\ep}=exp\left\{{\ep \over 2} [\gamma_E-\ln 4\pi]\right\}
\end{eqnarray}
is the spherical factor characteristic of $n$-dimensional regularisation.

The fact that $\hat a_s$ is independent of the choice of $\mu_R$ leads to
the following renormalisation group equation (RGE) for the coupling constant:
\begin{eqnarray}
\mu_R^2 {d \ln a_s(\mu_R^2) \over d \mu_R^2}
={\ep \over 2} + {1 \over a_s(\mu_R^2)}~ \beta(a_s(\mu_R^2))
\end{eqnarray}
where
\begin{eqnarray}
\beta(a_s(\mu_R^2)) &=& -a_s(\mu_R^2)~\mu_R^2 {d \ln Z(\mu_R^2) \over d \mu_R^2}
=-\sum_{i=0}^\infty a_s^{i+2}(\mu_R^2)~ \beta_i
\end{eqnarray}
The solution to the above equation is given by
\begin{eqnarray}
Z(\mu_R^2)= 1+ a_s(\mu_R^2) {2 \beta_0 \over \ep}
           + a_s^2(\mu_R^2) \Bigg({4 \beta_0^2 \over \ep^2 }+
                  {\beta_1 \over \ep} \Bigg)
           + a_s^3(\mu_R^2) \Bigg( {8 \beta_0^3 \over \ep^3}
                   +{14 \beta_0 \beta_1 \over 3 \ep^2}
                   +{2 \beta_2 \over 3 \ep}\Bigg)
\end{eqnarray}
The renormalisation constant $Z(\mu_R^2)$ relates the bare coupling constant 
$\hat a_s$ to the renormalised one $a_s(\mu_R^2)$ through
the eqn.(\ref{renas}).
The coefficients $\beta_0$ and $\beta_1$ are 
\begin{eqnarray}
\beta_0&=&{11 \over 3 } C_A - {4 \over 3 } T_F n_f
\nonumber \\
\beta_1&=&{34 \over 3 } C_A^2-4 T_F n_f C_F -{20 \over 3} T_F n_f C_A
\end{eqnarray}
where the color factors for $SU(N)$ QCD are given by
\begin{eqnarray}
C_A=N,\quad \quad \quad C_F={N^2-1 \over 2 N} , \quad \quad \quad 
T_F={1 \over 2}
\end{eqnarray}
and $n_f$ is the number of active flavours.  In the case of the Higgs
production, the number of active flavours is five because the
top degrees of freedom is integrated out in the large $m_{top}$ limit.

The factors $Z^I(\hat a_s,\mu_R^2,\mu^2,\ep)$ are the overall 
operator renormalisation constants.
For the vector current  $Z^q(\hat a_s,\mu_R^2,\mu^2)=1$,
but the gluon operator gets overall renormalisation 
\cite{Chetyrkin:1997un} given by
\begin{eqnarray}
Z^g(\hat a_s,\mu_R^2,\mu^2,\ep)&=&
1+\hat a_s \left({\mu_R^2 \over \mu^2}\right)^{\ep \over 2}
  S_{\ep} ~\Bigg[{2 \beta_0 \over \ep}\Bigg]
 +\hat a_s^2 \left({\mu_R^2 \over \mu^2}\right)^{\ep}
S_{\ep}^2 ~\Bigg[{2 \beta_1 \over \ep} \Bigg]
\nonumber\\[2ex]
&&  +\hat a_s^3 \left ({\mu_R^2 \over \mu^2}\right)^{3{\ep \over 2}}
S_{\ep}^3~ \Bigg[ 
           {1 \over \ep^2}\Big(-2 \beta_0 \beta_1  \Big)
           +{2 \beta_2 \over \ep}\Bigg]
\end{eqnarray}

The bare form factors $\hat F^I(\hat a_s,Q^2,\mu^2)$ 
(before performing overall renormalisation) 
of both fermionic and gluonic operators satisfy
the following integro differential equation that follows from the gauge 
as well as 
renormalisation group invariances \cite{Sudakov:1954sw,Mueller:1979ih,
Collins:1980ih,Sen:1981sd}.  In dimensional regularisation,
\begin{eqnarray}
Q^2{d \over dQ^2} \ln \hat {F^I}\left(\hat a_s,Q^2,\mu^2,\ep\right)&=&
{1 \over 2 }
\Bigg[K^I\left(\hat a_s,{\mu_R^2 \over \mu^2},\ep\right)  
+ G^I\left(\hat a_s,{Q^2 \over \mu_R^2},{\mu_R^2 \over \mu^2},\ep\right)
\Bigg]
\label{sud1}
\end{eqnarray}
where $K^I$ contains all the poles in $\ep$.  
On the other hand, $G^I$ collects rest of the terms that are 
finite as $\ep$ becomes zero.  
In other words $G^I$ contains only non-negative powers of $\ep$. 
Since $\hat F^I$ is RG invariant,  we find
\begin{eqnarray}
\mu_R^2 {d \over d\mu_R^2} 
K^I\Bigg(\hat a_s,{\mu_R^2 \over \mu^2},\ep\Bigg)=-A^I(a_s(\mu_R^2))
\nonumber\\[2ex]
\mu_R^2 {d \over d\mu_R^2} 
G^I\Bigg(\hat a_s,{Q^2\over \mu_R^2},
{\mu_R^2 \over \mu^2},\ep\Bigg)=A^I(a_s(\mu_R^2))
\end{eqnarray}
The quantities $A^I$ are the standard cusp anomalous dimensions and they are
expanded in powers of renormalised strong coupling constant $a_s(\mu_R^2)$ as
\begin{eqnarray}
A^I(\mu_R^2)=\sum_{i=1}^\infty a_s^{i}(\mu_R^2)~ A_i^I
\end{eqnarray} 
The total derivative is given by
\begin{eqnarray}
\mu_R^2 {d \over d\mu_R^2} = \mu_R^2 {\partial \over \partial \mu_R^2}
+{d a_s(\mu_R^2) \over d\mu_R^2} {\partial \over \partial a_s(\mu_R^2)}
\end{eqnarray}

The RGE of $K^I$ can be solved in powers of bare coupling constant
$\hat a_s$ as
\begin{eqnarray}
K^I\left(\hat a_s,{\mu_R^2\over \mu^2},\ep\right)
=\sum_{i=1}^\infty \hat a_s^i 
\left({\mu_R^2 \over \mu^2}\right)^{i {\ep \over 2}}S^i_{\ep}~ K^{I,(i)}(\ep)
\end{eqnarray}
where,
\begin{eqnarray}
K^{I,{1}}(\ep)&=& {1 \over \ep} \Bigg(- 2 A_1^I\Bigg)
\nonumber\\[2ex]
K^{I,{2}}(\ep)&=& {1 \over \ep^2} \Bigg(2 \beta_0 A_1^I\Bigg)
                 +{1 \over \ep}\Bigg(- A_2^I\Bigg)
\nonumber\\[2ex]
K^{I,{3}}(\ep)&=& {1 \over \ep^3} \Bigg(-{8 \over 3} \beta_0^2 A_1^I\Bigg)
                 +{1 \over \ep^2} \Bigg({2 \over 3} \beta_1 A_1^I
                        +{8 \over 3} \beta_0 A_2^I \Bigg)
                 +{1 \over \ep} \Bigg(-{2 \over 3} A_3^I \Bigg)
\nonumber\\[2ex]
K^{I,{4}}(\ep)&=& {1 \over \ep^4} \Bigg(4 \beta_0^3 A_1^I\Bigg)
                 +{1 \over \ep^3} \Bigg(-{8 \over 3} \beta_0 \beta_1 A_1^I 
                      - 6 \beta_0^2 A_2^I \Bigg)
\nonumber\\[2ex]
&&                 +{1 \over \ep^2} \Bigg({1 \over 3} \beta_2 A_1^I 
                       +\beta_1 A_2^I + 3 \beta_0 A_3^I \Bigg)
                   +{1 \over \ep} \Bigg(-{1 \over 2} A_4^I\Bigg)
\end{eqnarray}

Similarly RGE for $G^I$ can also be solved and the solution is found to be 
\begin{eqnarray}
G^I\left(\hat a_s,{Q^2 \over \mu_R^2},{\mu_R^2 \over \mu^2},\ep\right)
&=& G^I \left(a_s(\mu_R^2),{Q^2 \over \mu_R^2},\ep\right)
\nonumber\\[2ex]
&=&G^I\left(a_s(Q^2),1,\ep\right)+ \int_{Q^2 \over \mu_R^2}^1 
{d\lambda^2 \over \lambda^2} A^I\left(a_s(\lambda^2 \mu_R^2)\right)
\end{eqnarray} 
The integral in the above equation can be performed and it is found
to be
\begin{eqnarray}
\int_{Q^2 \over \mu_R^2}^1 {d \lambda^2 \over \lambda^2} 
A^I\left(a_s(\lambda^2 \mu_R^2)\right)
&=& \sum_{i=1}^\infty \hat a_s^i
\left({\mu_R^2 \over \mu^2}\right)^{i {\ep \over 2}}
\left[\left({Q^2 \over \mu_R^2}\right)^{i {\ep \over 2}}-1\right]
S^i_{\ep}~ K^{I,(i)}(\ep)
\end{eqnarray}
The finite function $G^I(a_s(Q^2),1,\ep)$ can also be expanded 
in powers of $a_s(Q^2)$ as
\begin{eqnarray}
G^I(a_s(Q^2),1,\ep)=\sum_{i=1}^\infty a_s^i(Q^2)~ G^{I}_i(\ep)
\end{eqnarray}
After substituting these solutions in the eqn.(\ref{sud1}) and performing the final
integration,  we obtain the following solution 
\begin{eqnarray}
\ln \hat F^I(\hat a_s,Q^2,\mu^2,\ep)
=\sum_{i=1}^\infty \hat a_s^i 
\left({Q^2 \over \mu^2}\right)^{i {\ep \over 2}}S^i_{\ep}~ \hat {\cal L}_F^{I,(i)}(\ep)
\end{eqnarray}
where
\begin{eqnarray}
\hat {\cal L}_ F^{I,(1)}&=&{1\over \ep^2} \Bigg(-2 A_1^I\Bigg) 
              +{1 \over \ep} \Bigg(G_1^I(\ep)\Bigg)
\nonumber\\[2ex]
\hat {\cal L}_ F^{I,(2)}&=&{1\over \ep^3} \Bigg(\beta_0 A_1^I\Bigg) 
                  +{1\over \ep^2} \Bigg(-{1 \over 2} A_2^I 
                  - \beta_0  G_1^I(\ep)\Bigg)
                  +{1 \over 2 \ep} G_2^I(\ep)
\nonumber\\[2ex]
\hat {\cal L}_ F^{I,(3)}&=& {1\over \ep^4} \Bigg(-{8 \over 9}\beta_0^2 A_1^I\Bigg) 
                  + {1\over \ep^3} \Bigg({2 \over 9} \beta_1 A_1^I 
                    +{8 \over 9} \beta_0 A_2^I +{4 \over 3} 
                     \beta_0^2 G_1^I(\ep)\Bigg) 
\nonumber\\[2ex]
&&                  +{1\over \ep^2} \Bigg(-{2 \over 9} A_3^I 
                   -{1 \over 3} \beta_1 G_1^I(\ep) 
                   -{4 \over 3}\beta_0 G_2^I(\ep)\Bigg)
                  +{1 \over \ep}\Bigg({1 \over 3} G_3^I(\ep)\Bigg)
\nonumber\\[2ex]
\hat {\cal  L}_F^{I,(4)}&=& {1\over \ep^5} \Bigg(\beta_0^3 A_1^I\Bigg) 
                  +{1 \over \ep^4} \Bigg(-{2 \over 3} \beta_0 \beta_1 A_1^I
                   -{3 \over 2}\beta_0^2 A_2^I -2 \beta_0^3 G_1^I(\ep)\Bigg)
\nonumber\\[2ex]
&&                  +{1 \over \ep^3} \Bigg({1 \over 12} \beta_2 A_1^I 
                     +{1 \over 4} \beta_1 A_2^I 
                    + {3 \over 4}\beta_0 A_3^I +{4 \over 3} 
                       \beta_0 \beta_1 G_1^I(\ep)
                    +3\beta_0^2 G_2^I(\ep)\Bigg)
\nonumber\\[2ex]
&&                  +{1 \over \ep^2} \Bigg(-{1 \over 8} A_4^I
                       -{1 \over 6} \beta_2 G_1^I(\ep)
                       -{1 \over 2} \beta_1 G_2^I(\ep) 
                    -{3\over 2} \beta_0 G_3^I(\ep)\Bigg)
                  +{1 \over \ep} \Bigg({1 \over 4} G_4^I(\ep)\Bigg)
\end{eqnarray}
The above result is in agreement with \cite{Moch:2005id},  
which was evaluated using various algorithms designed for solving 
nested sums.  The cusp anomalous dimensions $A^I_i$
and $G^{I}_i(\ep)$ are known upto 
order $a_s^3$.  The cusp anomalous dimensions are maximally non-abelien
and hence satisfy the following relation:
\begin{eqnarray}
A^q={ C_F \over C_A }~A^g  
\end{eqnarray}

The coefficients $G^{I}_i(\ep)$ can be found for
both $I=q$ and $I=g$ in \cite{Moch:2005tm} to the required accuracy in
$\ep$.  They satisfy 
\begin{eqnarray}
G^{I}_1(\ep)&=& 
2 (B^I_1 - \delta_{I,g} \beta_0) + f_1^I
        +\sum_{k=1}^\infty \ep^k  g^{~I,k}_1
\nonumber \\[2ex]
G^{I}_2(\ep)&=& 
2(B_2^I-2 \delta_{I,g} \beta_1) + f_2^I 
        -2 \beta_0  g^{~I,1}_1
        +\sum_{k=1}^\infty \ep^k  g^{~I,k}_2
\nonumber \\[2ex]
G^{I}_3(\ep)&=&
2 (B_3^I - 3\delta_{I,g} \beta_2) + f_3^I 
        -2 \beta_1  g^{~I,1}_1
-2 \beta_0 \Big(g^{~I,1}_2+2 \beta_0 g^{~I,2}_1\Big)
\nonumber \\[2ex]
&&        +\sum_{k=1}^\infty \ep^k g^{~I,k}_3
\end{eqnarray}
The constants $B_i^I$ are also 
known upto order $a_s^3$ thanks to the recent computation
of three loop anomalous dimensions/splitting functions 
\cite {Moch:2004pa,Vogt:2004mw}.

The constants $f_i^I$ are analogous to the cusp anomalous dimensions
$A_i^I$ that enter the form factors.  It was first noticed in 
\cite{Ravindran:2004mb}
that the single pole (in $\ep$) of the logarithm of form factors 
upto two loop level ($a_s^2$) can be
predicted due the presence of these constants $f_i^I$ because they are
found to be maximally non-abelien obeying the relation 
\begin{eqnarray}
f_i^q={C_F\over C_A} f_i^g
\end{eqnarray}
similar to $A_i^I$.  In \cite{Moch:2005tm},
this relation has been found to hold even at the three loop level.

The partonic cross sections $\hat \sigma^{sv}_I(z,q^2,\mu_R^2)$ is UV finite 
after the coupling constant and overall operator renormalisations are
performed using $Z(\mu_R^2)$ and $Z^I(\mu_R^2)$.  
But they still require mass factorisation in order to remove the collinear 
divergences:  
\begin{eqnarray}
\hat \sigma^{sv}_I(z,q^2,\mu_R^2,\ep)=
      \Gamma^T(z,\mu_F^2,\ep)\otimes 
       \Delta^{sv}_{I}\left(z,q^2,\mu_R^2,\mu_F^2\right) \otimes\Gamma(z,\mu_F^2,\ep)
\end{eqnarray}
with $\mu_F$ being the factorisation scale.
The resulting coefficient functions $\Delta^{sv}_{I}(z,q^2,\mu_R^2,\mu_F^2)$ are 
finite and free of collinear singularities. 
\begin{eqnarray}
\Delta^{sv}_{I}(z,q^2,\mu_R^2,\mu_F^2)=
\delta(1-z)+\sum_{i=1}^{\infty}a_s^i(\mu_R^2)~ \Delta_{I}^{sv,(i)}
\left(z,q^2,\mu_R^2,\mu_F^2\right)
\end{eqnarray}
The coefficient functions $\Delta_{I}^{sv,(i)}$ for $i=1,2$ are known(see 
\cite{Dawson:1990zj} to \cite{Ravindran:2003um}).  
The partial result for $\Delta_{I}^{sv,(3)}$ (i.e., all ${\cal D}_i$ 
except $\delta(1-z)$ are known for $i=3$) is also available(see \cite{Moch:2005ky}).  

The kernel $\Gamma(z,\mu_F^2,\ep)$ satisfies the following 
renormalisation group equation: 
\begin{eqnarray}
\mu_F^2 {d \over d\mu_F^2}\Gamma(z,\mu_F^2,\ep)={1 \over 2}  P
                         \left(z,\mu_F^2\right) 
                        \otimes \Gamma \left(z,\mu_F^2,\ep\right)
\end{eqnarray}
The $P(z,\mu_F^2)$ are well known Altarelli-Parisi splitting
functions(matrix valued) known upto three loop level \cite{Moch:2004pa,Vogt:2004mw}:
\begin{eqnarray}
P(z,\mu_F^2)=
\sum_{i=1}^{\infty}a_s^i(\mu_F^2) P^{(i-1)}(z)
\end{eqnarray}
The diagonal terms of splitting functions 
$P^{(i)}(z)$ have the following structure
\begin{eqnarray}
P^{(i)}_{II}(z) = 2\Bigg[ B^I_{i+1} \delta(1-z) 
                  + A^I_{i+1} {\cal D}_0\Bigg] + P_{reg,II}^{(i)}(z)
\end{eqnarray}
where $P_{reg,II}^{(i)}$ are regular when the argument takes
the kinematic limit(here $z \rightarrow 1$).
The RGE of the kernel
can be solved in dimensional regularisation
in powers of strong coupling constant.  Since we are interested
only in the soft plus virtual part of the cross section, only the diagonal parts
of the kernels contribute.  In the $\overline{MS}$ scheme, 
the kernel contains only poles in $\ep$.  Expanding the kernel in powers
of bare coupling $\hat a_s$,  
\begin{eqnarray}
\Gamma(z,\mu_F^2,\ep)=\delta(1-z)+\sum_{i=1}^\infty \hat a_s^i 
\left({\mu_F^2 \over \mu^2}\right)^{i {\ep \over 2}}S^i_{\ep} 
\Gamma^{(i)}(z,\ep)
\end{eqnarray}
we can solve the RGE for the kernel.  The solutions in the  
$\overline{MS}$ scheme are given by
\begin{eqnarray}
\Gamma_{II}^{(1)}(z,\ep)&=&{1 \over \ep} \Pzz 
\nonumber\\[2ex]
\Gamma_{II}^{(2)}(z,\ep)&=&
                   {1 \over \ep^2}\Bigg({1 \over 2} \Pzz \otimes \Pzz 
                       -\beta_0 \Pzz\Bigg)
                 +{1 \over \ep} \Bigg({1 \over 2} \Poz\Bigg)
\nonumber\\[2ex]
\Gamma_{II}^{(3)}(z,\ep)&=&
               {1 \over \ep^3}\Bigg(
              {4 \over 3} \beta_0^2 \Pzz -\beta_0 \Pzz \otimes \Pzz
\nonumber\\[2ex]
&&              +{1 \over 6} \Pzz \otimes \Pzz \otimes \Pzz
              \Bigg) +
              {1 \over \ep^2} \Bigg( {1 \over 2} \Pzz \otimes \Poz
\nonumber\\[2ex]
&&                -{1 \over 3} \beta_1 \Pzz -{4 \over 3} \beta_0 \Poz\Bigg)
              +{1 \over \ep} \Bigg({1 \over 3} \Ptz\Bigg)
\nonumber\\[2ex]
\Gamma_{II}^{(4)}(z,\ep)&=&
            {1 \over \ep^4}\Bigg(
            {1 \over 24} \Pzz \otimes \Pzz \otimes \Pzz\otimes \Pzz
\nonumber\\[2ex]
&&           -{1 \over 2} \beta_0 \Pzz \otimes \Pzz \otimes \Pzz
            +{11 \over 6} \beta_0^2 \Pzz \otimes \Pzz
\nonumber\\[2ex]
&&           -2 \beta_0^3 \Pzz
             \Bigg)
\nonumber\\[2ex]
&&         +{1 \over \ep^3}\Bigg(
           {1 \over 4} \Pzz \otimes \Pzz \otimes \Poz
          -{1 \over 3} \beta_1 \Pzz \otimes \Pzz
\nonumber\\[2ex]
&&          -{11 \over 6} \beta_0 \Pzz \Poz
          +{4 \over 3} \beta_0 \beta_1 \Pzz
          +3 \beta_0^2 \Poz
           \Bigg)
\nonumber\\[2ex]
&&        +{1 \over \ep^2}\Bigg(
         {1\over 3}\Pzz \otimes \Ptz
          +{1 \over 8} \Poz \otimes \Poz
          -{1 \over 6} \beta_2 \Pzz
\nonumber\\[2ex]
&&          -{1 \over 2} \beta_1 \Poz
           -{3 \over 2} \beta_0 \Ptz\Bigg)
          +{1 \over \ep} \Bigg({1 \over 4} \Pthz
          \Bigg)
\end{eqnarray}
It is now straightforward to obtain the soft distribution functions
$\Phi^I(\hat a_s,q^2,\mu^2,z)$ from the available results known upto three loop level
for the form factors $\hat F^I$, the kernels $\Gamma_{II}$ and the
coefficient functions $\Delta^{sv}_I$(the $\delta(1-z)$ function part of 
$\Delta^{sv,(3)}_I$ is still unknown).
The fact that $\Delta^{sv}_I$ are finite in the limit $\ep \rightarrow 0$ implies 
that the soft distribution functions have pole structure in $\ep$ similar to that
of $\hat F^I$ and $\Gamma_{II}$.  Also, $\Phi^I(\hat a_s,q^2,\mu^2,z)$ satisfy
the renormalisation group equation: 
\begin{eqnarray}
\mu_R^2 {d \over d\mu_R^2}\Phi^I(\hat a_s,q^2,\mu^2,z,\ep)=0
\end{eqnarray}
From the above observations, it is natural to expect that the soft
distribution functions also satisfy Sudakov type 
integro differential equation that 
the form factors $\hat F^I(Q^2)$ satisfy(see eqn.(\ref{sud1})).  Hence,   
\begin{eqnarray}
q^2 {d \over dq^2}\Phi^I(\hat a_s,q^2,\mu^2,z,\ep) = 
{1 \over 2 }
\Bigg[\overline K^{~I}\left(\hat a_s,{\mu_R^2 \over \mu^2},z,\ep\right)  
+ \overline G^{~I}\left(\hat a_s,{q^2 \over \mu_R^2},{\mu_R^2 \over \mu^2},z,\ep\right)
\Bigg]
\label{sud2}
\end{eqnarray}
where again $\overline K^{~I}$ contains all the singular terms and 
$\overline G^{~I}$
are finite functions of $\ep$.  The renormalisation group invariance leads
to 
\begin{eqnarray}
\mu_R^2 {d\over d\mu_R^2} \overline K^{~I}
\Bigg(\hat a_s, {\mu_R^2 \over \mu^2},z,\ep\Bigg)=
-\overline A^{~I}(a_s(\mu_R^2)) \delta(1-z)
\nonumber\\[2ex]
\mu_R^2 {d \over d\mu_R^2} \overline G^{~I}
\Bigg(\hat a_s,{q^2 \over \mu_R^2},{\mu_R^2 \over \mu^2},z,\ep\Bigg)
=\overline A^{~I}(a_s(\mu_R^2)) \delta(1-z)
\end{eqnarray}
If $\Phi^I(\hat a_s,q^2,\mu^2,z,\ep)$ have to contain the right poles 
to cancel the poles
coming from $\hat F^I$,$Z^I$ and $\Gamma_{II}$ in order to
make $\Delta^{sv}_I$ finite, then $\overline A^{~I}$ have to satisfy 
\begin{eqnarray}
\overline A^{~I}=-A^I
\end{eqnarray}
The above relation along with the renormalisation group invariance
implies that
\begin{eqnarray}
\overline G^{~I}\left(\hat a_s,{q^2 \over \mu_R^2},{\mu_R^2 \over \mu^2},z,\ep\right)
&=&\overline G^{~I} \left(a_s(\mu_R^2),{q^2 \over \mu_R^2},z,\ep\right)
\nonumber\\[2ex]
&=&\overline G^{~I}\left(a_s(q^2),1,z,\ep\right)
- \delta(1-z) \int_{q^2 \over \mu_R^2}^1 
{d\lambda^2 \over \lambda^2} A^I\left(a_s(\lambda^2 \mu_R^2)\right)
\end{eqnarray}

Now it is now straight forward to determine all 
$\overline G^{~I}(a_s(q^2),1,z,\ep)$
from the available informations.  The functions 
$\overline G^{~I}(a_s(q^2),1,z,\ep)$
can be expanding in powers of $a_s(q^2)$ as
\begin{eqnarray}
\overline G^{~I}(a_s(q^2),1,z,\ep)=\sum_{i=1}^{\infty} a_s^i(q^2) 
~\overline G^{~I}_i(z,\ep)
\end{eqnarray}
The solution to the eqn(\ref{sud2}) can be obtained in the way we obtained
$\ln \hat F^I(Q^2)$.  Expanding the soft distribution functions 
in powers of bare coupling $\hat a_s$ as 
\begin{eqnarray}
\Phi^I\left(\hat a_s,q^2,\mu^2,z,\ep\right)=\sum_{i=1}^\infty \hat a_s^i 
\left({q^2 \over \mu^2}\right)^{i {\ep \over 2}}S^i_{\ep} 
~\hat \Phi^{I,(i)}(z,\ep)
\end{eqnarray}
we find the solution:
\begin{eqnarray}
\hat \Phi^{I,(i)}(z,\ep) = \hat {\cal L}_F^{I,(i)}(\ep)
\Bigg(A^I\rightarrow -\delta(1-z)~A^I,~~
G^I(\ep) \rightarrow \overline G^{~I}(z,\ep)\Bigg)
\end{eqnarray}
The finite functions $\overline G^{~I}_i(z,\ep)$
can be obtained using the mass factorisation formula by
demanding the finiteness of the coefficient functions
$\Delta^{sv,(i)}_I$.  
The RG invariance of theses soft functions 
and the simple rescaling $q \rightarrow (1-z) q$
imply that the following expansion is also the solution to
the integro differential equation:
\begin{eqnarray}
\Phi^I(\hat a_s,q^2,\mu^2,z,\ep) &=& \Phi^I(\hat a_s,q^2 (1-z)^2,\mu^2,\ep)
\nonumber\\[2ex]
&=&\sum_{i=1}^\infty \hat a_s^i \left({q^2 (1-z)^2 \over \mu^2}\right)^{i 
{\ep \over 2}} S_{\ep}^i \left({i~ \ep \over 1-z} \right)\hat \phi^{~I,(i)}(\ep)
\end{eqnarray}
where
\begin{eqnarray}
\hat \phi^{I,(i)}(\ep)=\hat {\cal L}_F^{I,(i)}(\ep)
\Bigg( A^I \rightarrow - A^I, G^I(\ep) \rightarrow \overline {\cal G}^I(\ep)
\Bigg)
\end{eqnarray}
The $z$ independent constants 
$\overline {\cal G}^I(\ep)$ in $\hat \phi^{~I,(i)}(\ep)$ can be obtained 
using the form factors, mass factorisation kernels and coefficient
functions $\Delta^{sv,(i-1)}_I$ expanded in powers of $\ep$ to 
the desired accuracy.  This is achieved by comparing the poles 
as well as non-pole terms  
in $\ep$ of $\hat \phi^{~I,(i)}(\ep)$ with those coming from the
form factors, overall renormalisation constants and splitting functions
and the lower order $\Delta^{sv,(i-1)}_I$.  
We find
\begin{eqnarray}
\overline {\cal G}^{~I}_1(\ep)&=&-f_1^I+
\sum_{k=1}^\infty \ep^k \overline {\cal G}^{~I,(k)}_1
\nonumber\\[2ex]
\overline {\cal G}^{~I}_2(\ep)&=&-f_2^I
-2 \beta_0 \overline{\cal G}_1^{~I,(1)}
+\sum_{k=1}^\infty\ep^k  \overline {\cal G}^{~I,(k)}_2
\nonumber\\[2ex]
\overline {\cal G}^{~I}_3(\ep)&=&-f_3^I
-2 \beta_1 \overline{\cal G}_1^{~I,(1)}
-2 \beta_0 \left(\overline{\cal G}_2^{~I,(1)}
+2 \beta_0 \overline{\cal G}_1^{~I,(2)}\right)
+\sum_{k=1}^\infty \ep^k \overline {\cal G}^{~I,(k)}_3
\end{eqnarray}
with
\begin{eqnarray}
\overline{\cal G}^{I,(1)}_1&=& C_I~ \overline{\cal G}^{~(1)}_1
\nonumber\\[2ex]
&=&C_I~ \Big(-3 \zeta_2\Big)
\nonumber\\[2ex]
\overline{\cal G}^{I,(2)}_1&=&C_I~ \overline{\cal G}^{~(2)}_1
\nonumber\\[2ex]
&=& C_I~ \Bigg({7 \over 3}  \zeta_3\Bigg)
\nonumber\\[2ex]
\overline{\cal G}^{I,(1)}_2&=&C_I~ \overline{\cal G}^{~(1)}_2
\nonumber\\[2ex]
&=& C_I C_A~ \Bigg({2428 \over 81} -{469 \over 9} \zeta_2
              +4 \zeta_2^2 -{176 \over 3} \zeta_3\Bigg) 
\nonumber\\[2ex]
&&             +C_I n_f~ \Bigg(-{328 \over 81} + {70 \over 9} \zeta_2
                +{32 \over 3} \zeta_3 \Bigg) 
\end{eqnarray}
where $C_I=C_F$ for $I=q$(DY) and $C_I=C_A$ for $I=g$(Higgs).

Using such compensating
$\hat \phi^{I,(i)}(\ep)$ and the following expansion,  
\begin{eqnarray}
{1 \over 1-z} \big[(1-z)^2\big]^{i {\ep \over 2}}
={1 \over i \ep}\delta(1-z) 
+ \sum_{j=0}^{\infty} { (i \ep)^j \over j!} {\cal D}_j
\end{eqnarray}
we obtain $\overline G^{~I}_i(z,\ep)$ upto three loop level.
We find that the finite functions $\overline G^{~I}_i(z,\ep)$ have 
the following decomposition
in terms of cusp anomalous dimension $A^I_i$ and $f^I_i$ that appear
in the form factors:
\begin{eqnarray}
\overline G^{{~I}}_1&=& -f_1^I ~\delta(1-z)+2 A_1^I~ {\cal D}_0 
        +\sum_{k=1}^\infty \ep^k \overline g^{~I,k}_1(z)
\nonumber \\[2ex]
\overline G^{{~I}}_2&=& -f_2^I ~\delta(1-z)+2 A_2^I~ {\cal D}_0
        -2 \beta_0 ~\overline g^{~I,1}_1(z)
        +\sum_{k=1}^\infty \ep^k ~\overline g^{~I,k}_2(z)
\nonumber \\[2ex]
\overline G^{{~I}}_3&=& -f_3^I ~\delta(1-z)+2 A_3^I~ {\cal D}_0
        -2 \beta_1 ~\overline g^{~I,1}_1(z)
-2 \beta_0 \Big(\overline g^{~I,1}_2(z)+2 \beta_0 ~\overline g^{~I,2}_1(z)\Big)
\nonumber \\[2ex]
&&        +\sum_{k=1}^\infty \ep^k ~\overline g^{~I,k}_3(z)
\end{eqnarray}
where
\begin{eqnarray}
\overline g^{~I,1}_1&=&C_I \Bigg( 8 {\cal D}_1 -3 \zeta_2 \delta(1-z)\Bigg)
\nonumber\\[2ex]
\overline g^{~I,2}_1&=&C_I \Bigg( -3 \zeta_2 {\cal D}_0 +
              4 {\cal D}_2 +{7 \over 3} \zeta_3 \delta(1-z)\Bigg)
\nonumber\\[2ex]
\overline g^{~I,3}_1&=&C_I \Bigg( {7 \over 3} \zeta_3 {\cal D}_0 -
              3 \zeta_2 {\cal D}_1 
              +{4 \over 3} {\cal D}_3
             -{3 \over 16} \zeta_2^2 \delta(1-z)\Bigg)
\nonumber\\[2ex]
\overline g^{~I,1}_2&=&C_I C_A \Bigg( 
              \Bigg(-{1616 \over 27} +{242 \over 3} \zeta_2
               +56 \zeta_3 \Bigg) {\cal D}_0 
              +\Bigg({1072 \over 9} -32 \zeta_2\Bigg) {\cal D}_1 
              +\Big(-88\Big) {\cal D}_2 
\nonumber\\[2ex]
&&             +\Bigg({2428 \over 81} -{469 \over 9} \zeta_2
              +4 \zeta_2^2 -{176 \over 3} \zeta_3\Bigg) \delta(1-z)\Bigg)
            +C_I n_f \Bigg(\Bigg({224 \over 27} 
                    -{44 \over 3} \zeta_2\Bigg) {\cal D}_0
\nonumber\\[2ex]
&&               +\Bigg(-{160\over 9}\Bigg) {\cal D}_1
               +16 {\cal D}_2 
             +\Bigg(-{328 \over 81} + {70 \over 9} \zeta_2
                +{32 \over 3} \zeta_3 \Bigg) \delta(1-z)\Bigg)
              \Bigg)
\nonumber\\[2ex]
\overline g^{~I,2}_2&=&C_I C_A \Bigg( \Bigg(
            {4856 \over 81} - {938 \over 9} \zeta_2 
             +8 \zeta_2^2 -{1210 \over 9} \zeta_3
            \Bigg) {\cal D}_0
           +\Bigg(-{3232 \over 27}+ {550 \over 3} \zeta_2 
               +112 \zeta_3\Bigg) {\cal D}_1
\nonumber\\[2ex]
&&           +\Bigg({1072 \over 9} -32 \zeta_2\Bigg) {\cal D}_2
           +\Bigg(-{616 \over 9} \Bigg){\cal D}_3  \Bigg) 
           +C_I n_f \Bigg(
          \Bigg(-{656 \over 81} +{140 \over 9} \zeta_2 
                 +{220 \over 9} \zeta_3 \Bigg) {\cal D}_0
\nonumber\\[2ex]
&&          +\Bigg({448 \over 27} -{100 \over 3} \zeta_2\Bigg) {\cal D}_1
          +\Bigg( -{160 \over 9} \Bigg){\cal D}_2
          +\Bigg( {112 \over 9}\Bigg) {\cal D}_3\Bigg)
          \Bigg) 
\nonumber\\[2ex]
&&        +  \delta(1-z) \delta \overline g^{g,2}_2
\nonumber\\[2ex]
\overline g^{~I,1}_3&=&
C_I C_A^2 \Bigg(
   \Bigg(
  -{403861 \over 243}-176 \zeta_2 \zeta_3 
  +{71584 \over 27} \zeta_2
  -{5368 \over 15} \zeta_2^2 +{9272\over 3} \zeta_3
  -576 \zeta_5
   \Bigg){\cal D}_0
\nonumber\\[2ex]
&&  + \Bigg(
  {257140 \over 81} -{28696 \over 9} \zeta_2
 +{1056 \over 5} \zeta_2^2
 -{5632 \over 3} \zeta_3
   \Bigg){\cal D}_1
  + \Bigg(
  -{68752 \over 27} +{1760 \over 3} \zeta_2
   \Bigg){\cal D}_2
\nonumber\\[2ex]
&&  + \Bigg(
  {7744 \over 9}
   \Bigg){\cal D}_3
\Bigg)
+C_I C_A n_f\Bigg(
\Bigg(
{96482 \over 243} -{2452 \over 3} \zeta_2
+{1264 \over 15} \zeta_2^2
-{6536 \over 9} \zeta_3
\Bigg){\cal D}_0
\nonumber\\[2ex]
&&+\Bigg(
-{72008 \over 81}+{9056 \over 9} \zeta_2
+{448 \over 3} \zeta_3
\Bigg){\cal D}_1
+\Bigg(
{22400 \over 27}-{320 \over 3} \zeta_2
\Bigg){\cal D}_2
+\Bigg(
-{2816 \over 9}
\Bigg){\cal D}_3\Bigg)
\nonumber\\[2ex]
&&+C_I n_f^2\Bigg(
\Bigg(
-{4480 \over 243}+{1520 \over 27} \zeta_2
+{416 \over 9} \zeta_3
\Bigg){\cal D}_0
+\Bigg(
{4192 \over 81}-{736 \over 9} \zeta_2
\Bigg){\cal D}_1
\nonumber\\[2ex]
&&+\Bigg(
-{1600 \over 27} 
\Bigg){\cal D}_2
+\Bigg(
{256 \over 9} 
\Bigg){\cal D}_3
\Bigg)
+C_I C_F n_f\Bigg(
\Bigg(
{1711 \over 9} -60 \zeta_2 
-{96 \over 5}\zeta_2^2-{304 \over 3} \zeta_3
\Bigg){\cal D}_0
\nonumber\\[2ex]
&&+\Bigg(
-220 +192 \zeta_3
\Bigg){\cal D}_1
+\Bigg(
64
\Bigg){\cal D}_2
\Bigg)+\delta\overline g^{g,1}_3 \delta(1-z)
\end{eqnarray}
In the above equation $\delta \overline g^{g,2}_2,\delta\overline g^{g,1}_3$ 
are not known
because the full fixed order $N^3LO$ computation for the soft part of
the cross section is not available yet.  

With the available informations(ignoring
$\delta \overline g^{g,2}_2,\delta \overline g^{g,1}_3$), 
we find that the soft distribution functions
for DY and Higgs productions are maximally non-abelien: 
\begin{eqnarray}
\Phi^q(\hat a_s,q^2,\mu^2,z,\ep)={C_F \over C_A}~ 
\Phi^g(\hat a_s,q^2,\mu^2,z,\ep) 
\end{eqnarray}
upto three loop level.  At the cross section level($\Delta_I^{sv}$),
this property does not show up because of the form factors which
do not have this property.  The overall factors $C_F$ and $C_A$ ordinate
from the leading order contributions to the soft distribution functions.
Hence if you factor out this colour factor ($C_F$ for the DY and $C_A$ for 
the Higgs) we find that the soft distribution functions are universal.
\begin{eqnarray}
\Phi^I(\hat a_s,q^2,\mu^2,z,\ep)=C_I \Phi(\hat a_s,q^2,\mu^2,z,\ep)
\end{eqnarray}
The universality of the soft distribution functions can be understood
if you notice that the soft part of the cross section is always
independent of the spin, colour, flavour or any other quantum
numbers after factoring out the born level cross section. 
It depends only on the gauge interaction, here it is $SU(N)$. 
This universal property can be utilised to compute
soft part of the any new cross section where incoming particles
carry any spin,colour,flavour or other quantum numbers. 
For example, if we know $\Phi(\hat a_s,q^2,\mu^2,z,\ep)$ extracted
from the Drell-Yan production results, we can predict
the $\Delta^{sv}_g(z,q^2,\mu_R^2,\mu_F^2)$ for the Higgs production using
mass factorisation formula
provided we know the gluon form factor $F^g(\hat a_s,Q^2,\mu^2,\ep)$
and the overall renormalisation constant $Z^g(\hat a_s,\mu_R^2,\mu^2,\ep)$.

The soft plus virtual part of the cross section 
($\Delta^{sv}_I(z,q^2,\mu_R^2,\mu_F^2)$)
using mass factorisation formula is found to be
\begin{eqnarray}
\Delta^{sv}_I(z,q^2,\mu_R^2,\mu_F^2)={\cal C} \exp
\Bigg({\Psi^I(z,q^2,\mu_R^2,\mu_F^2,\ep)}\Bigg)\Bigg|_{\ep=0}
\end{eqnarray}
where $\Psi^I(z,q^2,\mu_R^2,\mu_F^2,\ep)$ is a finite distribution.  
The $\mu_R$ dependence comes from the coupling constant
and operator renormalisation:
\begin{eqnarray}
\Psi^I(z,q^2,\mu_R^2,\mu_F^2,\ep)&=&
\Bigg(
\ln \Big(Z^I(\hat a_s,\mu_R^2,\mu^2,\ep)\Big)^2 
+\ln \big|\hat F^I(\hat a_s,Q^2,\mu^2,\ep)\big|^2 
\Bigg)
\delta(1-z)
\nonumber\\[2ex]
&&+2 C_I \Phi(\hat a_s,q^2,\mu^2,z,\ep) 
-2~ {\cal C}\ln \Gamma_{II}(\hat a_s,\mu^2,\mu_F^2,z,\ep)
\end{eqnarray}
In the above equation "${\cal C} \ln$" means the "convolution ordered"
logarithm.  All the products of distributions 
in the logarithmic expansion are understood as Mellin convolutions.
The distribution 
$\Psi^I(z,q^2,\mu_R^2,\mu_F^2,\ep)$ is regular as $\ep\rightarrow 0$.  The
soft plus virtual cross section can be obtained by expanding 
$\Psi^I(z,q^2,\mu_R^2,\mu_F^2,\ep=0)$ as
\begin{eqnarray}
\Psi^I(z,q^2,\mu_F^2,\ep)
=\sum_{i=1} a_s^i(\mu_F^2) \Psi^{~I,(i)}(z,q^2,\mu_F^2) 
\end{eqnarray}
where we have set $\mu_R=\mu_F$ and expressing $a_s(\mu_F^2)$ in terms
of $a_s(\mu_R^2)$ is straightforward.
We find that the cross sections $\Delta^{sv}_I(z,q^2,\mu_F^2)$
can be obtained using
\begin{eqnarray}
\Delta^{sv,(0)}_I(z,q^2,\mu_F^2)&=&C_I \delta(1-z)
\nonumber\\[2ex]
\Delta^{sv,(1)}_I(z,q^2,\mu_F^2)&=&\Psi^{~I,(1)}(z,q^2,\mu_F^2)
\nonumber\\[2ex]
\Delta^{sv,(1)}_I(z,q^2,\mu_F^2)&=&\Psi^{~I,(2)}(z,q^2,\mu_F^2)
+{1 \over 2} \Psi^{~I,(1)}(z,q^2,\mu_F^2)
\otimes \Psi^{~I,(1)}(z,q^2,\mu_F^2)
\nonumber\\[2ex]
\Delta^{sv,(3)}_I(z,q^2,\mu_F^2)&=&\Psi^{~I,(3)}(z,q^2,\mu_F^2)+\Psi^{~I,(1)}
(z,q^2,\mu_F^2)\otimes \Psi^{~I,(2)}(z,q^2,\mu_F^2)
\nonumber\\[2ex]
&&+{1 \over 6} \Psi^{~I,(1)}(z,q^2,\mu_F^2) \otimes \Psi^{~I,(1)}(z,q^2,\mu_F^2)
\otimes \Psi^{~I,(1)}(z,q^2,\mu_F^2)
\end{eqnarray}
where
\begin{eqnarray}
\Psi^{~I,(1)}&=&
\Big((-2 \beta_0 \delta_{I,g}+ 2 B^I_1)\delta(1-z)
+2 A^I_1 {\cal D}_0 \Big)\ln\left({q^2 \over \mu_F^2}\right)
+\Big(3 \zeta_2 A^I_1 +2 g_1^{~I,1} 
\nonumber\\[2ex]
&&+ 2 C_I \overline{\cal G}_1^{~(1)}\Big)\delta(1-z)
+ \Big(4 A^I_1\Big){\cal D}_1
\nonumber\\[2ex]
\Psi^{~I,(2)}&=&
\Bigg[6 \zeta_2 \beta_0 B^I_1
-6 \zeta_2 \beta_0^2 \delta_{I,g}
+2 \beta_0 g_1^{~I,2}
+g_2^{~I,1}
+3 \zeta_2 A^I_2
+2 \beta_0 C_I \overline {\cal G}_1^{~(2)}
+C_I \overline {\cal G}_2^{~(1)}
\nonumber\\[2ex]
&&
+\Bigg(
-4 \beta_1 \delta_{I,g}
-2 \beta_0 g_1^{(1)}
+2 B^I_2
-3\zeta_2 \beta_0 A^I_1
-2 \beta_0 C_I \overline {\cal G}_1^{~(1)}\Bigg)
\ln\left({q^2 \over \mu_F^2}\right)
\nonumber\\[2ex]
&&+\Big(-\beta_0 B^I_1 
+\beta_0^2 \delta_{I,g}\Big) \ln^2\left({q^2 \over \mu_F^2}\right)
\Bigg]\delta(1-z)
+\Bigg[
-4 \beta_0 C_I \overline {\cal G}_1^{~(1)}
-2 f_2^I 
+2 A^I_2 \ln\left({q^2 \over \mu_F^2}\right)
\nonumber\\[2ex]
&&-\beta_0 A^I_1 \ln^2\left({q^2 \over \mu_F^2}\right)
\Bigg] {\cal D}_0
+\Bigg[4 A^I_2 
-4 \beta_0 A^I_1 \ln\left({q^2 \over \mu_F^2}\right)
\Bigg]{\cal D}_1
+\Bigg[-4 \beta_0 A^I_1 \Bigg]{\cal D}_2
\nonumber\\[2ex]
                     \Psi^{~I,(3)}&=&
\Bigg[-30 \zeta_2 \beta_0 \beta_1\delta_{I,g}
+12\zeta_2 \beta_0 B^I_2
-12 \zeta_2 \beta_0^2 g_1^{~I,(1)}
+ 6 \zeta_2 \beta_1 B^I_1
+{4 \over 3} \beta_0 g_2^{~I,(2)}
+{8 \over 3} \beta_0^2 g_1^{~I,(3)}
\nonumber\\[2ex]
&&+{4 \over 3} \beta_1 g_1^{~I,(2)}
+{2 \over 3} g_3^{~I,(1)}
+3 \zeta_2 A^I_3
-3 \zeta_2^2 \beta_0^2 A^I_1
+{4 \over 3} \beta_0 C_I \overline {\cal G}_2^{(2)}
+{8 \over 3} \beta_0^2 C_I \overline {\cal G}_1^{~(3)}
\nonumber\\[2ex]
&&
+{4 \over 3} \beta_1 C_I \overline {\cal G}_1^{~(2)}
+{2 \over 3} C_I \overline {\cal G}_3^{~(1)}
+6 \zeta_2 \beta_0 f^I_2
+
\Bigg(
12 \zeta_2 \beta_0^3\delta_{I,g}
-6 \beta_2\delta_{I,g}
+2 B^I_3
-12 \zeta_2 \beta_0^2 B^I_1
\nonumber\\[2ex]
&&-2 \beta_0 g_2^{~I,(1)}
-4 \beta_0^2 g_1^{~I,(2)}
-2 \beta_1 g_1^{~I,(1)}
-6 \zeta_2\beta_0 A^I_2
-3 \zeta_2 \beta_1 A_1^I
-2 \beta_0 C_I \overline {\cal G}_2^{~(1)}
\nonumber\\[2ex]
&&-4 \beta_0^2 C_I \overline {\cal G}_1^{~(2)}
-2 \beta_1 C_I \overline {\cal G}_1^{~(1)}
\Bigg)\ln\left({q^2 \over \mu_F^2}\right)
+
\Bigg(
5 \beta_0 \beta_1 \delta_{I,g}
-2 \beta_0 B^I_2
-\beta_1 B^I_1
+2 \beta_0^2 g_1^{~I,(1)}
\nonumber\\[2ex]
&&+3 \zeta_2 \beta_0^2 A^I_1
+2 \beta_0^2 C_I \overline {\cal G}_1^{~(1)}
\Bigg)\ln^2 \left({q^2 \over \mu_F^2}\right)
+\Bigg({2 \over 3 } \beta_0^2 B^I_1
-{2 \over 3}\beta_0^3 \delta_{I,g}\Bigg)
\log^3\left({q^2 \over \mu_F^2}\right)
\Bigg]
\delta(1-z)
\nonumber\\[2ex]
&&+\Bigg[
-4 \beta_0 C_I \overline {\cal G}_2^{~(1)}
-8 \beta_0^2 C_I \overline {\cal G}_1^{~(2)}
-4 \beta_1 C_I \overline {\cal G}_1^{~(1)}
-2 f_3^I
+\Bigg(4 \beta_0 f_2^I 
+8 \beta_0^2 C_I \overline {\cal G}_1^{~(1)}
\nonumber\\[2ex]
&&+2 A^I_3 \Bigg) \ln\left({q^2 \over \mu_F^2}\right)
+\Bigg(-2 \beta_0 A^I_2 -\beta_1 A^I_1 \Bigg) 
\ln^2 \left({q^2 \over \mu_F^2}\right)
+\Bigg({2 \over 3} \beta_0^2 A^I_1 \Bigg) 
\ln^3 \left({q^2 \over \mu_F^2}\right)\Bigg]{\cal D}_0
\nonumber\\[2ex]
&&+\Bigg[
8 \beta_0 f_2^I 
+16 \beta_0^2 C_I \overline {\cal G}_1^{~(1)}
+4 A_3^I
+\Big(-8 \beta_2 A_2^I 
 -4 \beta_1 A^I_1\Big)\ln\left({q^2 \over \mu_F^2}\right)
\nonumber\\[2ex]
&&+\Big(4 \beta_0^2 A^I_1\Big) \ln^2 \left({q^2 \over \mu_F^2}\right)
\Bigg] {\cal D}_1 
+\Bigg[
-8 \beta_0 A^I_2 
-4 \beta_1 A^I_1
+\Big(8 \beta_0^2 A^I_1\Big) \ln\left({q^2 \over \mu_F^2}\right)
\Bigg]{\cal D}_2
\nonumber\\[2ex]
&&+\Bigg[
\Bigg({16 \over 3} \beta_0^2 A^I_1 \Bigg)
\Bigg]{\cal D}_3
\end{eqnarray}
Notice that $\Psi^{~I,(1)}$ and $\Psi^{~I,(2)}$ are
completely known.  Using our $\Psi^{~I,(1)}$,
and $\Psi^{~I,(2)}$ we could successfully reproduce
soft plus virtual cross section $\Delta_g^{sv,(i)}(z,q^2,\mu_F^2)$ 
(see \cite{Harlander:2001is,Catani:2001ic})
($i=1,2$) of the Higgs production
from that of DY \cite{Matsuura:1988sm}
and vice versa.  To compute $\Delta_I^{sv,(3)}(z,q^2,\mu_F^2)$
(equivalently $\Psi^{~I,(3)}$) we need to know
$\overline {\cal G}_2^{~I,(2)}$ and $\overline {\cal G}_3^{~I,(1)}$ 
either from DY or Higgs production because these constants are
maximally non-abelien.  Notice that these constants 
appear only in the coefficient
of $\delta(1-z)$ part of $\Psi^{~I,(3)}$.  Since the coefficients of
${\cal D}_i$($i=0,1,2,3$) in $\Psi^{~I,(3)}$ do not depend on these unknown
constants $\overline {\cal G}_3^{~I,(1)}$ and
$\overline {\cal G}_2^{~I,(2)}$, we can predict these
coefficients(say for the Higgs production)
by using the universal soft distribution function extracted
from a process(say DY), the three loop
form factors and the renormalisation constants.   
Our prediction agrees with the partial $N^3LO$ soft plus virtual
results \cite{Moch:2005ky} for DY and Higgs productions.

With these available informations one can also determine all the quantities
upto $N^3LL$ level in the threshold resummation.  To do this, we first
recollect that the soft distribution function is renormalisation group
invariant.  Its UV divergence can be removed by 
the coupling constant renormalisation.  This introduces
a renormalisation scale $\mu_R$ which is arbitrary to all orders
in perturbation theory.  In order to compute
various quantities in the threshold resummation
formula from the soft distribution function, we choose $\mu_R=\mu_F$. 
With this choice, one can express
the soft distribution function as a sum of pole and finite
parts in $\ep$ as $\ep \rightarrow 0$, that is
\begin{eqnarray}
\Phi^I\left(a_s(\mu_F^2),{q^2 \over \mu_F^2},z,\ep\right)
=\Phi^I_{pole}\Bigg(a_s(\mu_F^2),{q^2 \over \mu_F^2},z,{1 \over \ep}\Bigg)
               +\Phi^I_{fin}\Bigg(a_s(\mu_F^2),{q^2\over \mu_F^2},z,\ep\Bigg) 
\end{eqnarray}
With this decomposition,
it is now straightforward to identify the finite part 
$\Phi^I_{fin}$
with the threshold resummation formula as
\begin{eqnarray}
2 \int_0^1 dz ~z^{N-1} \Phi^I_{fin}\Bigg(a_s(\mu_F^2),
               {q^2 \over \mu_F^2},z,\ep=0\Bigg)
&=&\int_0^1 dz {z^{N-1}- 1\over 1-z} 
  \Bigg[ D^I\Big(a_s\Big(q^2(1-z)^2\Big)\Big) 
\nonumber\\[2ex]
&&  + 2 \int_{\mu_F^2}^{q^2 (1-z)^2} {d \lambda^2 \over \lambda^2} 
      A^I\Big(a_s(\lambda^2)\Big)  \Bigg]
\nonumber\\[2ex]
&&     +H^I_S\Bigg(a_s(\mu_F^2),{q^2 \over \mu_F^2}\Bigg) 
\end{eqnarray}
where the subscript $S$ in $H^I_S$ indicates that it 
comes from the soft part of the cross section.  The
remaining contribution comes from the form factor.
$D^I(a_s(q^2(1-z)^2))$ can be expanded in powers of bare coupling
constant $\hat a_s$ as follows:
\begin{eqnarray}
D^I\Big(a_s\Big(q^2 (1-z)^2\Big)\Big)=\sum_{i=1}^\infty 
\hat a_s^i \Bigg({q^2 (1-z)^2 \over \mu^2}\Bigg)^{i {\ep \over 2}}
S_{\ep}^i~ \hat D^{I,(i)}(\ep)
\end{eqnarray}
The finiteness of $D^I$ after coupling constant renormalisation
demands that it satisfies the following expansion in $\ep$: 
\begin{eqnarray}
\hat D^{I,(i)}(\ep)=\sum_{j=1-i}^\infty \hat d^{~I,(i)}_{j}  \ep^{j}
\end{eqnarray}
Using RG invariance, the coefficients of negative powers of 
$\ep$ can be evaluated as
\begin{eqnarray}
\hat d^{~g,(2)}_{-1}&=& -2 \beta_0 ~\hat d^{~g,(1)}_0
\nonumber\\[2ex]
\hat d^{~g,(3)}_{-2}&=& 4 \beta_0^2 ~\hat d^{~g,(1)}_0
\nonumber\\[2ex]
\hat d^{~g,(3)}_{-1}&=& -4 \beta_0 ~\hat d^{~g,(2)}_0
                -4 \beta_0^2 ~\hat d^{~g,(1)}_1 
             -\beta_1 ~\hat d^{~g,(1)}_0
\end{eqnarray}
We find that for non-negative powers of $\ep$($j\ge 0$),
\begin{eqnarray}
\hat d_j^{~I,(1)}&=&2 \overline g_1^{~I,j+1}\Big|_{{\cal D}_0}
\nonumber\\[2ex]
\hat d_j^{~I,(2)}&=&\Bigg(\overline g_2^{~I,j+1}
               -2 \beta_0 ~\overline g_1^{~I,j+2} \Bigg)\Big|_{{\cal D}_0}
\nonumber\\[2ex]
\hat d_j^{~I,(3)}&=&\Bigg({2 \over 3} ~\overline g_3^{~I,j+1}
                  -{8 \over 3} \beta_0 ~\overline g_2^{~I,j+2}
                  -{2 \over 3} \beta_1 ~\overline g_1^{~I,j+2}
                  +{8 \over 3} \beta_0^2 
                    ~\overline g_1^{~I,j+3}\Bigg)\Big|_{{\cal D}_0}
\end{eqnarray}
Using the above equations, we find explicitly
\begin{eqnarray}
\hat d^{~g,(1)}_0&=&0
\nonumber\\[2ex]
\hat d^{~g,(1)}_1&=&C_A \Big(- 6 \zeta_2\Big)
\nonumber\\[2ex]
\hat d^{~g,(1)}_2&=&C_A \Bigg({14 \over 3} \zeta_3 \Bigg)
\nonumber\\[2ex]
\hat d^{~g,(2)}_0&=&C_A^2 \Bigg(
         -{1616 \over 27 } +{308 \over 3} \zeta_2 +56 \zeta_3 
         \Bigg)
       + C_A n_f \Bigg(
         {224 \over 27} -{56 \over 3} \zeta_2 \Bigg)
\nonumber\\[2ex]
\hat d^{~g,(2)}_1&=&C_A^2 \Bigg({4856 \over81} -{938 \over 9} \zeta_2
         +8 \zeta_2^2 -{1364 \over 9} \zeta_3 \Bigg)
\nonumber\\[2ex]
&&        +C_A n_f\Bigg(
          -{656 \over 81} +{140 \over 9} \zeta_2
           +{248 \over 9} \zeta_3 \Bigg)
\nonumber\\[2ex]
\hat d^{~g,(3)}_0&=&C_A^3 \Bigg(
         -{1235050 \over 729}-{352 \over 3} \zeta_2 \zeta_3
         +{227548 \over 81} \zeta_2 -{1584 \over 5} \zeta_2^2
         +{10376 \over 3} \zeta_3 -384 \zeta_5
           \Bigg)
\nonumber\\[2ex]
&&       +C_A^2 n_f \Bigg(
        {328388 \over 729} -{72004 \over 81} \zeta_2
         +{352 \over 5} \zeta_2^2 -{26800 \over 27} \zeta_3
        \Bigg)
\nonumber\\[2ex]
&&       +C_A C_F n_f \Bigg(
         {3422 \over 27} -44 \zeta_2 -{64 \over 5} \zeta_2^2
         -{608 \over 9} \zeta_3
              \Bigg)
\nonumber\\[2ex]
&&       +C_A n_f^2 \Bigg(
        -{19456 \over 729} +{1760 \over 27} \zeta_2
        +{2080 \over 27} \zeta_3
         \Bigg)
\end{eqnarray}
The coefficients $\hat d^{~q,(i)}_{j}$ for the DY can be obtained using
\begin{eqnarray}
\hat d^{~q,(i)}_{j}={C_F \over C_A} ~\hat d^{~g,(i)}_{j}
\end{eqnarray}
because the soft distributions functions are maximally non-abelien.
Also, the coefficients of $a_s^i (q^2) {\cal D}_0$ in the soft distribution function
$\Phi^I_{fin}$ are related to
the coefficients $D_i^I$ that appear in threshold resummation 
formula.  Hence it is straightforward to 
obtain $D^I_i$ from the soft distribution function $\Phi^I_{fin}$. 
We find that $D^I_i$ are related to $\hat d^{~I,(i)}_k$ and hence
$\overline g_i^{~I,k}$ as follows:
\begin{eqnarray}
D^I_1&=& \hat d^{~I,(1)}_0 =2 ~\overline g_1^{~I,1} \Big|_{{\cal D}_0}
\nonumber\\[2ex]
&=&2 \overline{\cal G}_1^I(\ep=0)
\nonumber\\[2ex]
D^I_2&=&\hat d^{~I,(2)}_0+2 \beta_0 ~\hat d^{~I,(1)}_1
=\Bigg(\overline g_2^{~I,1}
        +2 \beta_0 ~\overline g_1^{~I,2}\Bigg)\Big|_{{\cal D}_0}
\nonumber\\[2ex]
&=&2 \overline{\cal G}_2^I(\ep=0)
\nonumber\\[2ex]
D^I_3 &=&\hat d^{~I,(3)}_0
          +4 \beta_0 ~\hat d^{~I,(2)}_1
        +\beta_1 ~\hat d^{~I,(1)}_1
       +4 \beta_0^2 ~\hat d^{~I,(1)}_2
\nonumber\\[2ex]
&=&\Bigg({2 \over 3} ~\overline g_3^{~I,1}
       +{4 \over 3} \beta_0 ~\overline g_2^{~I,2}
       +{4 \over 3} \beta_1 ~\overline g_1^{~I,2}
       +{8 \over 3} \beta_0^2 ~\overline g_1^{~I,3}
        \Bigg)\Big|_{{\cal D}_0}
\nonumber\\[2ex]
&=&2 \overline{\cal G}_3^I(\ep=0)
\end{eqnarray}
From the available informations on three loop results,
we find that the following result holds 
\begin{eqnarray}
D_i^I&=&2 ~\overline {\cal G}_i^I(\ep=0)
\quad \quad \quad \quad i=1,2,3
\end{eqnarray}
The fact that $D^I_i$ can be expressed entirely in terms of 
$\overline G^{~I}_i(z,\ep)$
(i.e., in terms of $\overline g^{~I,k}_i$ or 
$\overline {\cal G}_i^I(\ep=0)$) 
which are maximally non-abelien,
implies that $D^I_i$ are also maximally non-abelien.   Hence,
\begin{eqnarray}
D^q_i={C_F \over C_A}~ D^g_i 
\end{eqnarray}
with 
\begin{eqnarray}
D^g_1  & =&0
\nonumber\\[2ex]
D^g_2  & =&
   C_A^2 \left( - {1616\over 27} + {176\over 3}\,\zeta_2 + 56\,\zeta_3 \right)
   +  C_A n_f \left( {224\over 27} - {32\over 3}\,\zeta_2 \right)
\nonumber\\[2ex]
D^g_3  & =&
      C_A^3 \Bigg( - {594058 \over 729} + {98224 \over 81}\,\zeta_2
                   + {40144 \over 27}\,\zeta_3 - {2992 \over 15}\,\zeta_2^2
                   - {352 \over 3}\,\zeta_2\zeta_3 - 384\,\zeta_5 \Bigg)
\nonumber \\[2ex]
&&    + C_A^2 n_f  \Bigg( {125252 \over 729} - {29392 \over 81}\,\zeta_2
                   - {2480 \over 9}\,\zeta_3 + {736 \over 15}\,\zeta_2^2 \Bigg)
\nonumber \\[2ex]
&&    + C_A C_F n_f \Bigg( {3422 \over 27} - 32\,\zeta_2
                   - {608 \over 9}\,\zeta_3 - {64 \over 5}\,\zeta_2^2 \Bigg)
    + C_A n_f^2 \Bigg( - {3712 \over 729} + {640 \over 27} \,\zeta_2
                   + {320 \over 27}\,\zeta_3 \Bigg)
\end{eqnarray}
The above results are in agreement 
with \cite{Moch:2005ky,Laenen:2005uz,Idilbi:2005ni}. 
We also find the resummation exponents
$D^I_i$ can be extracted by using the following
relations:
\begin{eqnarray}
D^I_1&=&  
       \Delta_I^{sv,(1)}\Big|_{{\cal D}_0}
\nonumber\\[2ex]
D^I_2&=&  
        \Bigg(\Delta_I^{sv,(2)} -{1 \over 2} \Delta_I^{sv,(1)}\otimes
        \Delta_I^{sv,(1)}\Bigg) \Big|_{{\cal D}_0}
\nonumber\\[2ex]
D^I_3&=&
        \Bigg(\Delta_I^{sv,(3)} -\Delta_I^{sv,(1)} \otimes \Delta_I^{sv,(2)}
       +{1 \over 3} \Delta_I^{sv,(1)} \otimes \Delta_I^{sv,(1)} \otimes 
\Delta_I^{sv,(1)}\Bigg) \Big|_{{\cal D}_0}
\end{eqnarray}
where $\Delta_I^{sv,(i)}$ in the above are computed at 
the scale $\mu_F^2=\mu_R^2=q^2$.
From the following convolution identity\cite{vanNeerven:2001pe} 
upto irrelevant regular terms(denoted by $\cdot \cdot \cdot$)
\begin{eqnarray}
{\cal D}_i \otimes {\cal D}_j = d_{ij} \delta(1-z) 
+\sum_{l=0}^{i+j+1} c_{ij,l} {\cal D}_l
+\cdot \cdot \cdot
\end{eqnarray}
it is interesting to notice that in order to obtain 
$D^I_{i}$, it is sufficient
to know the coefficients of all ${\cal D}_{l}$ ($l=l_{max}$ to $0$)
(that means, we need not know the information on the coefficient of 
$\delta(1-z)$ function and the regular part of $\Delta_I^{sv,(i)}$) 
and the complete soft information of $\Delta_I^{sv,(i-1)}$(i.e., the
coefficients of all ${\cal D}_i$ and $\delta(1-z)$ are needed).

Finally, the coefficient of $\delta(1-z)$ in the resummation
formula can be obtained from $\Phi^I_{fin}$ by defining
the coupling constant at the scale $\mu_F^2$.  The result is
\begin{eqnarray}
H^I_S\left(a_s(\mu_F^2),{q^2 \over \mu_F^2}\right)
=\sum_{i=1}^\infty a_s^i(\mu_F^2) H^{I}_{S,i}
\end{eqnarray}
where 
\begin{eqnarray}
H^{g}_{S,1}&=&-3 \zeta_2 + \ln^2\left({q^2 \over \mu_F^2}\right)
\nonumber\\[2ex]
H^{g}_{S,2}&=& C_A^2 \Bigg(-{164 \over 81} +{35 \over 9} \zeta_2
                    +{34 \over 9} \zeta_3
+\Bigg(-{8 \over 3} \zeta_2 +{56 \over 27}  
     \Bigg)\ln\left({q^2 \over \mu_F^2}\right)
\nonumber\\[2ex]&&
     -{10 \over 9}\ln^2\left({q^2 \over \mu_F^2}\right)
     +{2 \over 9}\ln^3\left({q^2 \over \mu_F^2}\right)\Bigg)
+C_A n_f \Bigg({1214 \over 81} -{469 \over 18} \zeta_2
                    +2 \zeta_2^2-{187 \over 9}  \zeta_3
\nonumber\\[2ex]&&
+\Bigg({44 \over 3} \zeta_2 +14 \zeta_3 -
     {404 \over 27}\Bigg)\ln\left({q^2 \over \mu_F^2}\right)
+\Bigg(-2 \zeta_2 +{67 \over 9}\Bigg)
\ln^2\left({q^2 \over \mu_F^2}\right)
\nonumber\\[2ex]&&
-{11\over 9}
\ln^3\left({q^2 \over \mu_F^2}\right)\Bigg)
\end{eqnarray}
and 
\begin{eqnarray}
H^{q,(i)}_S={C_F \over C_A}~ H^{g,(i)}_S
\end{eqnarray}
The remaining contribution to the exponent comes from the
the finite part of form factor.

We conclude our discussion on this subject with a
brief discussion on the corresponding soft as well as 
jet distribution functions that appear in deep inelastic scattering.
The soft plus virtual coefficient function $c^{sv}_{I,2}(Q^2,z)$ 
that appear in the hadronic structure function $F_2$ can be expressed as
\begin{eqnarray}
{\cal C}\ln c_{I,2}^{sv}(Q^2,z)&=&
\Bigg(
\ln \Big(Z^I(\hat a_s,\mu_R^2,\mu^2,\ep)\Big)^2 
+\ln \big|\hat F^I(\hat a_s,Q^2,\mu^2,\ep)\big|^2 
\Bigg)
\delta(1-z)
\nonumber\\[2ex]
&&+ 2 \Phi^I_{SJ}(\hat a_s,Q^2,\mu^2,z,\ep) 
-{\cal C}\ln \Gamma_{II}(\hat a_s,\mu^2,\mu_F^2,z,\ep)
\end{eqnarray}
where $\Phi^I_{SJ}(\hat a_s,Q^2,\mu^2,z,\ep)$ is sum of
soft and jet distribution functions.  We find that this
soft plus jet distribution function also satisfies Sudakov
type integro differential equation (see eqn.(\ref{sud2})) 
which can be solved
in the same way we solved soft distribution functions.
We find that this soft plus jet distribution function
$\Phi^I_{SJ}$ can be expressed as
\begin{eqnarray}
\Phi^I_{SJ}(\hat a_s,Q^2,\mu^2,z,\ep) &
=& \Phi^I_{SJ}(\hat a_s,Q^2 (1-z),\mu^2,\ep)
\nonumber\\[2ex]
&=&\sum_{i=1}^\infty \hat a_s^i \left({Q^2 (1-z) \over \mu^2}\right)^{i 
{\ep \over 2}} S_{\ep}^i \left({i~ \ep \over 2(1-z)} \right)
\hat \xi^{~I,(i)}(\ep)
\end{eqnarray}
where
\begin{eqnarray}
\hat \xi^{I,(i)}(\ep)=\hat {\cal L}_F^{I,(i)}(\ep)
\Bigg( A^I \rightarrow - A^I, G^I(\ep) \rightarrow \widetilde {\cal G}^I(\ep)
\Bigg)
\end{eqnarray}
We find that the constants 
$\widetilde {\cal G}^I(\ep)$ have the following expansion in terms
of $B_i^I$, $f^I_i$ and the $\ep$ dependent part of 
lower order coefficient functions. 
\begin{eqnarray}
\widetilde {\cal G}^{~q}_1(\ep)&=&-(B_1^q+f_1^q)+
\sum_{k=1}^\infty \ep^k \widetilde {\cal G}^{~q,(k)}_1
\nonumber\\[2ex]
\widetilde {\cal G}^{~q}_2(\ep)&=&-(B_2^q+f_2^q)
-2 \beta_0 \widetilde{\cal G}_1^{~q,(1)}
+\sum_{k=1}^\infty\ep^k  \widetilde {\cal G}^{~q,(k)}_2
\nonumber\\[2ex]
\widetilde {\cal G}^{~q}_3(\ep)&=&-(B_3^q+f_3^q)
-2 \beta_1 \widetilde{\cal G}_1^{~q,(1)}
-2 \beta_0 \left(\widetilde{\cal G}_2^{~q,(1)}
+2 \beta_0 \widetilde{\cal G}_1^{~q,(2)}\right)
+\sum_{k=1}^\infty \ep^k \widetilde {\cal G}^{~q,(k)}_3
\end{eqnarray}
The $z$ independent constants $\widetilde {\cal G}^{q,(k)}_i$
are computed using the coefficient functions $c^{sv}_{q,2}(z,Q^2)$
known upto three loop level \cite{vanNeerven:1991nn,Moch:2005ba}.
Recollect that the three loop form factors were obtained from these
coefficient functions by demanding the finiteness of
the partonic cross sections after mass factorisation 
and also notice that the method used there is very different
from the method presented in this paper.  We obtain
\begin{eqnarray}
\widetilde{\cal G}^{q,(1)}_1&=&
C_F~ \Big({7 \over 2}-3 \zeta_2\Big)
\nonumber\\[2ex]
\widetilde{\cal G}^{q,(2)}_1&=&
C_F~ \Bigg(-{7 \over 2} +{9 \over 8} \zeta_2 +{7 \over 3}  \zeta_3\Bigg)
\nonumber\\[2ex]
\widetilde{\cal G}^{q,(1)}_2&=&
C_F^2 \Bigg(
{9 \over 8} -{41 \over 2}\zeta_2 
         +{82 \over 5} \zeta_2^2 -6 \zeta_3\Bigg)
\nonumber\\[2ex]
&&+C_F C_A~ \Bigg({69761 \over 648}-{1961 \over 36} \zeta_2
   -{17 \over 5} \zeta_2^2 -40 \zeta_3\Bigg)
\nonumber\\[2ex]
&&+ C_F n_f\Bigg(
-{5569 \over 324} +{163 \over 18} \zeta_2 +4 \zeta_3\Bigg)          
\end{eqnarray}
Using the following decomposition,
\begin{eqnarray}
\Phi^I_{SJ}\left(a_s(\mu_F^2),{Q^2 \over \mu_F^2},z,\ep\right)
=\Phi^I_{SJ,pole}\Bigg(a_s(\mu_F^2),{Q^2 \over \mu_F^2},z,{1 \over \ep}\Bigg)
               +\Phi^I_{SJ,fin}\Bigg(a_s(\mu_F^2),{Q^2\over \mu_F^2},z,\ep\Bigg)
\end{eqnarray}
it is now straightforward to identify the finite part
$\Phi^I_{SJ,fin}$ with the DIS threshold resummation formula as
\begin{eqnarray}
2 \int_0^1 dz ~z^{N-1} \Phi^I_{SJ,fin}\Bigg(a_s(\mu_F^2),
               {Q^2 \over \mu_F^2},z,\ep=0\Bigg)
&=&\int_0^1 dz {z^{N-1}- 1\over 1-z}
  \Bigg[ B_{DIS}^I\Big(a_s\Big(Q^2(1-z)\Big)\Big)
\nonumber\\[2ex]
&&  + \int_{\mu_F^2}^{Q^2 (1-z)} {d \lambda^2 \over \lambda^2}
      A^I\Big(a_s(\lambda^2)\Big)  \Bigg]
\nonumber\\[2ex]
&&     +H^I_{SJ,S}\Bigg(a_s(\mu_F^2),{Q^2 \over \mu_F^2}\Bigg)
\end{eqnarray}
Using the above equation, we find that 
the resummation constants $B^q_{DIS,i}$ 
satisfy the following relation
\begin{eqnarray}
B^q_{DIS,i}= \widetilde {\cal G}^q_i(\ep=0) \quad \quad \quad i=1,2,3
\end{eqnarray}
The resulting $B^q_{DIS,i}$s agree with those given in
\cite{Moch:2005ba}.  

To summarise, we have extracted
the soft distribution function $\Phi^I$ using mass factorisation
formula for both Drell-Yan as well as Higgs productions within the
framework of perturbative QCD.
This is possible now thanks to various three loop results available
for the form factors and splitting functions.
The $\Phi^I$ is known completely upto two loop level.  
Except the soft bremsstrahlung contributions proportional to $\delta(1-z)$
(at three loop level), all the other soft terms$({\cal D}_i)$ 
are known for the soft distribution functions 
$\Phi^I$ upto three loop level.  
We have also shown that the soft distribution functions satisfy 
Sudakov type integro-differential equation that 
the quark and gluon form factors satisfy.  
We found that they are process independent.  
In other words, knowing the soft distribution function 
of the Drell-Yan process, one can obtain
the same for the Higgs production by simply multiplying 
the colour factor combination $C_A/C_F$.  
Hence, unlike the cross sections $\Delta^{sv}_I$,
these soft distribution functions are maximally non-abelien.
Notice that the form factors have this property only
at the pole level.  
Using the universal soft distribution function extracted
from DY production, we could reproduce the 
soft plus virtual cross section for the Higgs production
using the gluon form factor and its renormalisation constant.   
We have also shown how these soft
distribution functions are related to
threshold resummation formula.   We have extracted various coefficients
appearing in threshold resummation formula upto
three loop level using the soft distribution function.  We have also
discussed the DIS soft and jet distribution functions in the present
context.

Acknowledgments:  The author would like to thank S. Moch and A. Vogt
for providing the constants ${\cal S}_i$ of DY that are computed
in their paper \cite{Moch:2005ky}.

%------------------------------
\end{document}